\documentclass{article}
\usepackage{amsmath,amsthm}
\usepackage{graphicx,epstopdf,epsfig,multirow,epic,bm}

\normalsize \rm
\begin{document}

\begin{center}
{\Large  \textbf { Structural properties and average tapping time on scale-free graphs with smallest diameter }}\\[12pt]
{\large Fei Ma$^{a,}$\footnote{~The author's E-mail: mafei123987@163.com. },\quad  Ping Wang$^{b,c,d,}$\footnote{~The corresponding author's E-mail: pwang@pku.edu.cn.} }\\[6pt]
{\footnotesize $^{a}$ School of Electronics Engineering and Computer Science, Peking University, Beijing 100871, China\\
$^{b}$ National Engineering Research Center for Software Engineering, Peking University, Beijing, China\\
$^{c}$ School of Software and Microelectronics, Peking University, Beijing  102600, China\\
$^{d}$ Key Laboratory of High Confidence Software Technologies (PKU), Ministry of Education, Beijing, China}\\[12pt]
\end{center}

\begin{quote}
\textbf{Abstract:} In this paper, we propose a class of graphs $G^{\star}(m,t)$ and first study some structural properties, such as, average degree, on them. The results show that (1) graphs $G^{\star}(m,t)$ have density feature because of their average degrees proportional to time step $t$ not to a constant in the large graph size limit, (2) graphs $G^{\star}(m,t)$ obey the power-law distribution with exponent equal to $2$, which is rarely found in most previous scale-free models, (3) graphs $G^{\star}(m,t)$ display small-world property in terms of ultra-small diameter and higher clustering coefficient, and (4) graphs $G^{\star}(m,t)$ possess disassortative structure with respect to Pearson correlation coefficient smaller than zero. In addition, we consider the trapping problem on the proposed graphs $G^{\star}(m,t)$ and then find that they all have more optimal trapping efficiency by means of their own average trapping time achieving the theoretical lower bound, a phenomenon that is seldom observed in existing scale-free models. We conduct extensive simulations that are consistent with our theoretical analysis.\\

\textbf{Keywords:} Scale-free graph, Structural properties, Trapping problem, Average trapping time, Maximum independent set. \\

\end{quote}

\vskip 1cm

The past two decades have witnessed an upsurge of complex network study \cite{Newman-2018}-\cite{Staudt-2016}. One of significant reasons for this is that various complex systems, both synthetic and real-world, can be naturally represented as complex networks, for instance, the World Wide Web, protein-protein interaction network, metabolic network, as well as friendship network, and so forth \cite{Newman-2018}. In general, the underlying structure of a complex network is a graph, denoted by $G(V,E)$ in this paper. Here, symbol $V$ is vertex set and $E$ represents edge set, and then $|V|$ denotes vertex number of graph $G(V,E)$, also called the order of graph, and $|E|$ is the total number of edges in graph $G(V,E)$, also defined as size of graph. At the same time, all graphs discussed in this paper are simple and connected, i.e., no loops and multi-edges. Hereafter, the terms network and graph will be used indistinctly.

A large number of empirical observations on complex networks in the last have shown that there are many interesting properties popularly found in various networks, such as, small-world phenomena \cite{ Watts-1998} and scale-free feature \cite{Albert-1999-1}. Since then, in order to better understand these properties, numerous networked models have been developed. As such, this further triggers the research of graphs themselves.

There are two main directions in graph research at present. The one is to study some structural properties on graph, for example, diameter and clustering coefficient (defined in detail later). Here, in the jargon of graph theory, diameter of a graph $G(V,E)$, denoted by $D$, is the maximum over distances of all possible vertex pairs. For a pair of vertices $u$ and $v$, distance between them, denoted by $d_{uv}$, is the edge number of any shorted path joining vertex $u$ and $v$. As known, graph with diameter equal to one must be the complete one. The structural properties related to the complete graph have been well studied over the past decades mainly because of their own specific structure. Interested reader can refer to \cite{Bondy-2008} for more detail. On the other hand, for a graph with diameter $2$, some commonly-focused structural properties will are significantly different from that of the complete graph with the same vertex set as will be shown shortly. Motivated by this, in this paper, we will generate a class of graphs with diameter exactly equivalent to $2$ and then study many other structural properties including degree distribution on them. Meanwhile, we precisely determine some closed-form solutions of quantities corresponding to those structural properties. One of them is that our graphs have scale-free feature because their degree distribution follow  power-law with exponent $2$ in form, which is rare in the context of complex network.

The other is to learn about dynamics taking placing on graphs, such as, random walks \cite{Sarma-2015}, percolation phase transition \cite{Colomer-de-Simon-2014} and synchronization \cite{Ling-2019}, and understand how the underlying structure of graph affects the dynamical behavior under consideration. To make further progress, this can in turn serve as instruments for designing more available graphs. Here, we focus mainly on a special kind of random walks, i.e., the trapping problem, on the proposed graph of great interest, at least, with diameter $2$. As above, by using a quantity, average trapping time ($ATT$) (defined in detail later), related to the trapping problem, we compare the exact solution to $ATT$ on our graph to that of complete graph and find that there are remarkable difference between them. In another words, a slight change of diameter results in a tremendous influence on determining formulas of $ATT$. For example, when placing traps at some vertices of our scale-free graph, the analytical value for $ATT$ can be asymptotically equal to $1$. Meantime, such an assignment of traps uncovers a unique maximum independent set on our graph. This reveals that it is considerably necessary to probe the nature of some interesting graphs like models introduced herein.

The rest of this paper is organized as follows. In Section 2, we provide a generative framework for constructing graphs with diameter $2$. Next, we study some structural properties on the proposed graphs and exactly derive the solutions corresponding to some related parameters in Section 3. Following the preceding section, the goal of Section 4 is to consider the trapping problem on the presented graphs and determine closed-form formula for average trapping time. Finally, we conclude our work in Section 5.

\section{Construction}

This section aims to provide a framework for generating graph $G^{\star}(m,t)$. As will be stated shortly, the candidate graph $G^{\star}(m,t)$ is in fact built up upon a hierarchical graph $G(m,t)$ in our prior work \cite{MF-2019-1}. First, let us construct graph $G(m,t)$ in an iterative manner.
\begin{itemize}
\item At $t=0$, the initial graph $G(m,0)$ is a star graph with $m$ ($\geq2$) leaf vertices, as illustrated in the top-left panel of Fig.1.
\item At $t=1$, the young graph $G(m,1)$ is generated by connecting a new vertex to each leaf vertex in number $m$ of graphs $G_{i}(m,0)$ ($i=1,2,\dots,m$) that are in essence duplications of graph $G(m,0)$. Such a process is depicted in the left-most panel in line 2 of Fig.1. For convenience, let that newly added vertex be at the layer $0$ of graph $G(m,1)$, all the centers from number $m$ of duplications $G_{i}(m,0)$ at the layer $1$ and the remaining of vertices in graph $G(m,1)$ at the layer $2$. After adopting such a rearrangement of vertices in graph $G(m,1)$, the hierarchical structure can be easily observed.

\item At $t\geq 2$, the newborn graph $G(m,t)$ can be obtained from the preceding graph $G(m,t-1)$ in a similar manner as in the second time step. Additionally, the only adjustment is to let the label of layer in graph $G(m,t-1)$ increase by a factor $1$.

The procedure above can be implemented iteratively until one anticipated graph $G(m,t)$ is obtained. After that, our graph $G^{\star}(m,t)$, the object discussed in this paper, may be established in an additional manipulation.

\item For a graph $G(m,t)$ ($t\geq1$), the corresponding graph $G^{\star}(m,t)$ is constructed from graph $G(m,t)$ by connecting that vertex at the layer $0$ to each vertex at the layer $i$ ($1\leq i\leq t$). An illustrative example is shown in the bottom-right panel of Fig.1. Hereafter, that vertex at the layer $0$ in graph $G^{\star}(m,t)$ is regarded as the hub, denoted by $\mathcal{H}(t)$, since it has the greatest degree.
\end{itemize}

\begin{figure}
\centering
  \includegraphics[height=7cm]{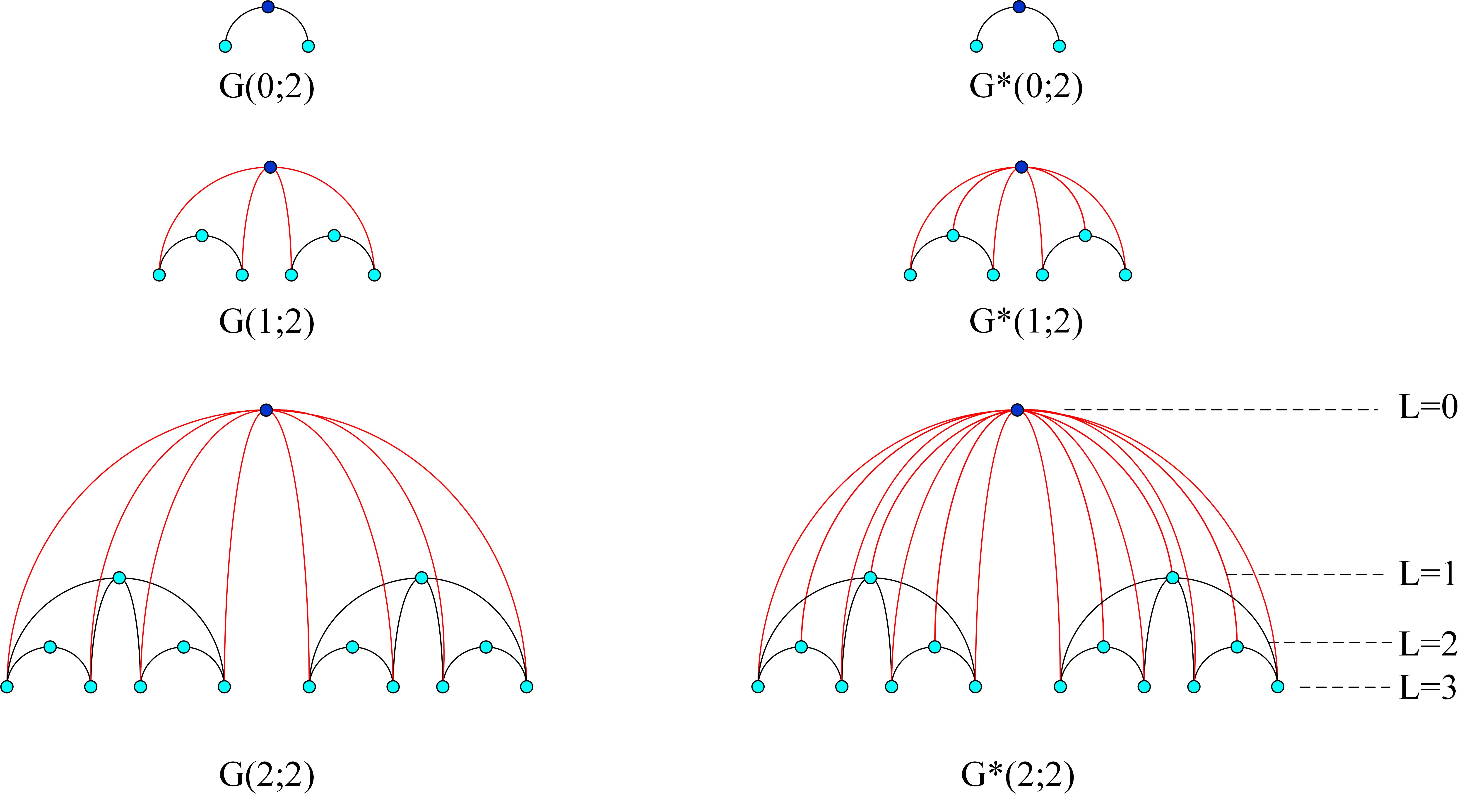}\\
{\small Fig.1. The diagram of the first three generations of graphs $G(m,t)$ and $G^{\star}(m,t)$ where parameter $m$ equals $2$.  }
\end{figure}

Obviously, the diameter of graph $G^{\star}(m,t)$ is exactly equal to $2$. And then, according to the hierarchial structure, the order $|\mathcal{V}^{\star}_{m}(t)|$ and size $|\mathcal{E}^{\star}_{m}(t)|$ of graph $G^{\star}(m,t)$ follow

\begin{subequations}
\label{eq:whole}
\begin{eqnarray}
|\mathcal{V}^{\star}_{m}(t)|=|\mathcal{V}_{m}(t)|=m|\mathcal{V}_{m}(t-1)|+1,\label{subeq:MF-2-1}
\end{eqnarray}
\begin{equation}
|\mathcal{E}^{\star}_{m}(t)|=|\mathcal{E}_{m}(t)|+\frac{m^{t+1}-m}{m-1}, \quad |\mathcal{E}_{m}(t)|=m|\mathcal{E}_{m}(t-1)|+m^{t+1}, \label{subeq:MF-2-2}
\end{equation}
\end{subequations}
where $|\mathcal{V}_{m}(t)|$ and $|\mathcal{E}_{m}(t)|$ denote, respectively, the order and size of graph $G(m,t)$. Using the initial conditions $|\mathcal{V}_{m}(0)|=m+1$ and $|\mathcal{E}_{m}(0)|=m$, the precise formulas of $|\mathcal{V}^{\star}_{m}(t)|$ and $|\mathcal{E}^{\star}_{m}(t)|$ can be derived in a recursive fashion, as follows

\begin{equation}\label{eqa:MF-2-3}
|\mathcal{V}^{\star}_{m}(t)|=\frac{m^{t+2}-1}{m-1}, \quad |\mathcal{E}^{\star}_{m}(t)|=\frac{(t+1)m^{t+2}-t\times m^{t+1}-m}{m-1}.
\end{equation}

So far, we have accomplished the generative construction of graph $G^{\star}(m,t)$. In the following, as tried in the context of complex networked models \cite{Wang-2019}-\cite{Chen-2007}, some commonly studied structural parameters  on graph $G^{\star}(m,t)$ will be considered in more detail, such as average degree. Besides that, a specific type of discrete-time unbiased random walks, namely, the trapping problem, is also studied on graph $G^{\star}(m,t)$ and we then precisely calculate the solution to average trapping time, an important quantity that is closely related to the trapping problem.

\section{Structural properties}

It is convention to understand the topological structure of a deterministic graph $G(V,E)$ in question by studying some structural parameters of considerable interest. Some of them, for instance, average degree, can be precisely calculated by means of some simple arithmetics. In this section, we will focus mainly on determining several structural parameters, for instance, average degree, vertex degree distribution, clustering coefficient and Pearson correlation coefficient, of graph $G^{\star}(m,t)$ built.

\subsection{Average degree}

Mathematically, the average degree $\langle k\rangle$ of a given graph $G(V,E)$ can be defined as the ratio of the summation over degrees $k_{i}$ of all vertices in graph $G(V,E)$ and the order $|V|$, namely,

\begin{equation}\label{eqa:MF-3-1-0}
\langle k\rangle=\frac{\sum_{i\in V}k_{i}}{|V|}=\frac{2|E|}{|V|}.
\end{equation}

In the study of complex networks, the average degree is a simple yet useful measure that indicates whether a network is sparse or not. Roundly speaking, in the limit of large graph size, a graph $G(V,E)$ is considered sparse if its average degree is approximately equivalent to an invariable. Otherwise, it is dense. By definition, we can have the following proposition

\textbf{Proposition 1} The average degree $\langle k^{\star}(m;t)\rangle$ of the proposed graph $G^{\star}(m,t)$ is given by

\begin{equation}\label{eqa:MF-3-1-1}
\langle k^{\star}(m;t)\rangle=O\left(\left(2-\frac{1}{m}\right)t\right),
\end{equation}
which is a variable varying linearly with time step $t$, suggesting that graph $G^{\star}(m,t)$ has density feature.

\textbf{\emph{ Proof}} Inserting the results from Eq.(\ref{eqa:MF-2-3}) into Eq.(\ref{eqa:MF-3-1-0}) outputs

\begin{equation}\label{eqa:MF-3-1-2}
\begin{aligned}\langle k^{\star}(m;t)\rangle&=2\frac{(t+1)m^{t+2}-t\times m^{t+1}-m}{m^{t+2}-1}\\
&\approx 2(t+1)-\frac{1}{m}t.
\end{aligned}
\end{equation}
This completes the proof of Proposition 1.

As opposed to some previous graphs in \cite{Wang-2019,Albert-2001}, graph $G^{\star}(m,t)$ shows density feature. So, such a structural property can be viewed as one virtue of our graph $G^{\star}(m,t)$. It is worth noting that the novel generation way mentioned in Section 2 leads graph $G^{\star}(m,t)$ to be dense. Besides that, the graph $G^{\star}(m,t)$ itself has many other interesting features in comparison with graphs in \cite{Ma-2019,Ravasz-2003} as will be reported shortly.

\subsection{Degree distribution}

In the last, as one fundamental structure feature related to a graph $G(V,E)$, degree sequence has been widely studied in various scientific community \cite{Bondy-2008}, especially in graph theory. In the jargon of graph theory, the degree sequence of a graph $G(V,E)$ consisting of vertices $v_{1}, v_{2},\dots, v_{|V|}$ can defined as ($k_{1}, k_{2},\dots, k_{|V|}$). Apparently, this is a set of discrete values. It is well known that for a given graph $G(V,E)$, the degree sequence can be equivalent to the degree distribution. Here, the probability for at random choosing a vertex with degree $k_{j}$ in graph $G(V,E)$ is defined to be

$$P(k_{j})=\frac{N_{k_{j}}}{|V|}$$
where $N_{k_{j}}$ is the total number of vertices whose degrees are exactly equal to $k_{j}$.

Indeed, such a convenient transformation from degree sequence into degree distribution has helped us to uncover some fundamental properties on a large number of graphs, for example, the degree distribution of ER-graph obeys Poisson one in form, implying that the degrees of a great body of vertices are close to that average value $\sum_{k=k_{min}}^{k_{max}}kP(k)$ \cite{Albert-2016}. In addition, in the past two decades, a lot of empirical observations on complex networks have shown that most networks follow the power-law degree distribution

$$P(k_{j})\sim k_{j}^{-\gamma}, \quad 2< \gamma\leq 3$$
meaning that a small fraction of vertices possess a large fraction of connections. Graphs of such type always exhibit heterogeneous structure significantly distinct from the underlying structure of ER-graph mentioned above.

In what follows, we will determine which type of degree distribution the presented graph $G^{\star}(m,t)$ obeys. Before starting our calculations, we need to introduce the accumulate degree distribution, a well-used technique that is suitable to calculating degree distribution of deterministic graph $G(V,E)$,

\begin{equation}\label{eqa:MF-3-2-0}
P_{cum}(k_{j})=\frac{\sum_{k\geq k_{j}}N_{k}}{|V|}
\end{equation}
where $N_{k}$ has the same definition as above.

\textbf{Proposition 2} In an overwhelm range, the degree distribution of graph $G^{\star}(m,t)$ follows

\begin{equation}\label{eqa:MF-3-2-1}
P(k)\sim k^{-\gamma}, \quad \gamma=2.
\end{equation}

\textbf{\emph{ Proof}} This suffices to divide all the vertices of graph $G^{\star}(m,t)$ into different classes according to vertex degree in order to consolidate Proposition 2. Taking into consideration the development of graph $G^{\star}(m,t)$, we can capture a list as below

\begin{center}
\begin{tabular}{c|c|c|c|c|c|c|c|cccc}
  \hline
  $k_{m}(t_{i},t)$ & $\frac{m^{t+2}-1}{m-1}-1$ & $m^{t}+1$ & $...$ & $m^{t_{i}}+1$ & $...$ & $m^{2}+1$ & $m+1$ & $t+1$ \\
   \hline
 $N_{m}(t_{i},t)$ & $1$ & $m$ & $...$ & $m^{t-t_{i}+1}$ & $...$ & $m^{t-1}$ & $m^{t}$& $m^{t+1}$ \\
  \hline
\end{tabular}
\end{center}
in which symbol $N_{m}(t_{i},t)$ denotes the total number of vertices with degree equal to $k_{m}(t_{i},t)$. It is obvious to see that the degree value $t+1$ must be smaller than $m^{j}+1$ for some parameter $j$. Without loss of generality, suppose that $t<m^{j_{t}}$ and then it is clear to find that most degree values fall into the range $\Gamma_{m}(t)$ from $m^{j_{t}}+1$ to $m^{t}+1$, an overwhelm range that has been demonstrated in Proposition 2.

Now, for a given degree value $k=m^{s}+1$ in range $\Gamma_{m}(t)$, the accumulate degree distribution can by definition in Eq.(\ref{eqa:MF-3-2-0}) be expressed as

\begin{equation}\label{eqa:MF-3-2-2}
P_{cum}(k)=\frac{\sum_{k_{i}\geq k}N_{k}}{|\mathcal{V}^{\star}_{m}(t)|}=\frac{1+\sum_{i=s}^{t}m^{t-i+1}}{|\mathcal{V}^{\star}_{m}(t)|}\sim m^{-s}.
\end{equation}

Meantime, one can solve for $s$ from equality $k=m^{s}+1$ to get $s=\ln(k-1)/\ln m$. And then, plugging that value for $s$ into Eq.(\ref{eqa:MF-3-2-2}) produces

\begin{equation}\label{eqa:MF-3-2-3}
P_{cum}(k)\sim k^{-1}.
\end{equation}

At present, taking differential on the both sides of Eq.(\ref{eqa:MF-3-2-3}), we have

\begin{equation}\label{eqa:MF-3-2-4}
P(k)\sim k^{-2},
\end{equation}
a result that is completely the same as that in Proposition 2, revealing that Proposition 2 holds true.

A similar approach to the above can be adopted to determine the degree distribution of those vertices with degrees smaller than $t+1$, so we omit it here.

Token together, in the large graph size limit, our graph $G^{\star}(m,t)$ obeys the power-law degree distribution and thus has scale-free feature, a character that is prevalent in a wide variety of complex networked models \cite{Newman-2018}. On the other hand, it is worth noting that theoretical models with power-law exponent $2$ are rarely reported in the literature of complex networked models. Hence, the lights shed by generating graph $G^{\star}(m,t)$ may be helpful to create many other scale-free models with density feature in the future.

\subsection{Clustering coefficient}

As another fundamental structure parameter, clustering coefficient $\langle C\rangle$ of a graph $G(V,E)$ under consideration, defined as the average over all vertices $v$, has attracted much attention in the past few years \cite{Newman-2018,Albert-2016}. Here, the clustering coefficient $c_{v}$ for a vertex $v$ in graph $G(V,E)$ is commonly regarded as the ratio of number $n_{v}$ of edges actually existing in the neighbor set of vertex $v$ and all possible edges number $k_{v}(k_{v}-1)/2$ where $k_{v}$ is the degree of vertex $v$, i.e., $c_{v}=\frac{2n_{v}}{k_{v}(k_{v}-1)}$. Thus, for a given graph $G(V,E)$, the clustering coefficient $\langle C\rangle$ is

\begin{equation}\label{eqa:MF-3-3-0}
\langle C\rangle=\frac{\sum_{v\in V}c_{v}}{|V|}.
\end{equation}

By definition, it is evident to see that clustering coefficient $\langle C\rangle$ measures the level of ``cluster" of a graph in question. The higher the value for $\langle C\rangle$ is, the more clustered graph is. As such, this parameter has been used in complex networks study as an index for determining whether a networked model considered is small-world or not \cite{Definition-2020-2}.

To evaluate the ``cluster" phenomena of our graph $G^{\star}(m,t)$ as a whole, the next task is to first derive the exact formula of clustering coefficient for each vertex in $G^{\star}(m,t)$ and then average those values over all vertices. Doing so leads to the following proposition.

\textbf{Proposition 3} The clustering coefficient $\langle C(m;t)\rangle$ of graph $G^{\star}(m,t)$ is equal to

\begin{equation}\label{eqa:MF-3-3-1}
\langle C(m;t)\rangle= \frac{\frac{2m-2}{m^{t+2}-2m+1}+\sum_{i=1}^{t}\frac{2m^{t-i+1}}{m^{i}+1}+\frac{2m^{t+1}}{t+1}}{\frac{m^{t+2}-1}{m-1}}.
\end{equation}

\textbf{\emph{ Proof}} According to the hierarchical structure of graph $G^{\star}(m,t)$, we can observe that all vertices lying at the same layer have a local structure in common. Hence, determining the clustering coefficients for those vertices at the same layer reduces to determining the clustering coefficient for an arbitrary vertex at that layer. The next discussions will be in progress by means of evaluating the clustering coefficients for vertices from number $t+2$ of different layers.

\begin{itemize}
\item For the hub, the degree value is equal to $\frac{m^{t+2}-1}{m-1}-1$ and the number of edges actually existing in its neighbor set is $m|V^{\star}(m,t-1)|=m\times\frac{m^{t+1}-1}{m-1}$. So, the clustering coefficient $c_{0}$ satisfies
    \begin{equation}\label{eqa:MF-3-3-2}
c_{0}= \frac{2}{\frac{m^{t+2}-1}{m-1}-2}.
\end{equation}

\item For an arbitrary vertex at the layer $i$ ($1\leq i\leq t$), its degree is equivalent to $m^{t+1-i}+1$ and the number of already existing edges between vertices in its neighbor set equals $m^{t+1-i}$. By definition, one has the clustering coefficient $c_{i}$

   \begin{equation}\label{eqa:MF-3-3-3}
c_{i}= \frac{2}{m^{t+1-i}+1}.
\end{equation}

\item Finally, for an arbitrary vertex from the bottom-most layer, it is obvious to find the degree value to be $t+1$ and there are number $t$ of edges allocated between vertices in its neighbor set. As previously, one may calculate the clustering coefficient $c_{t+1}$ to obtain
    \begin{equation}\label{eqa:MF-3-3-4}
c_{t+1}= \frac{2}{t+1}.
\end{equation}

\end{itemize}

Now, armed with both Eq.(\ref{eqa:MF-3-3-0}) and the results from Eqs.(\ref{eqa:MF-3-3-2})-(\ref{eqa:MF-3-3-4}), one is able to capture the closed-form solution for clustering coefficient $\langle C(m;t)\rangle$ of graph $G^{\star}(m,t)$

    \begin{equation}\label{eqa:MF-3-3-5}
\langle C(m;t)\rangle=\frac{\frac{2}{\frac{m^{t+2}-1}{m-1}-2}+2\sum_{i=1}^{t}\frac{m^{t-i+1}}{m^{i}+1}+2\frac{m^{t+1}}{t+1}}{\frac{m^{t+2}-1}{m-1}}.
\end{equation}
This is consistent with that in Eq.(\ref{eqa:MF-3-3-1}), indicating that Proposition 3 is valid.

As a case study, assume that parameter $m$ is equal to $2$, the value for $\langle C(2;t)\rangle$ can be asymptotically close to $\sum_{i=1}^{t}\frac{1}{2^{i}(2^{i}+1)}+\frac{1}{t+1}>0.217$ in the limit of large $t$. This suggests that graph $G^{\star}(2,t)$ has a non-zero clustering coefficient. Similarly, for many other smaller parameters $m$, graph $G^{\star}(m,t)$ always displays ``cluster" property because its clustering coefficient $\langle C(m;t)\rangle$ is strictly larger than $0$.

\subsection{Pearson correlation coefficient}

The previous three subsections focus mainly on characterizing properties closely correlated to vertex itself, while this subsection aims at investigating the feature of edge by virtue of its two endpoint degrees. In theory, such an investigation is to quantify tendency of connections taking place between vertices in a graph $G(V,E)$ indirectly. For instance, in social network research, a large volume of empirical observations imply a fact that it is mostly likely for two persons with the similar status to make friend and such a relationship between them is often stable overtime. On the other hand, in biological networks, it is popular to observe connections between vertices with distinct importance. In the language of mathematics, the two types of networks above can be naturally transformed into graphs considered here. After that, ``status" and ``importance" can be simply defined as the degree (or weight) of corresponding vertex. In order to plausibly evaluate phenomena of such kind on graphs $G(V,E)$, Newman provided in \cite{Newman-2002} a measurement, called Pearson correlation coefficient $r$, that is referred to as

\begin{equation}\label{eqa:MF-3-4-1}
r=\frac{|E|^{-1}\sum\limits_{e_{ij}\in E} k_{i}k_{j}-\left[|E|^{-1}\sum\limits_{e_{ij}\in E} \frac{1}{2}(k_{i}+k_{j})\right]^{2}}{|E|^{-1}\sum\limits_{e_{ij}\in E} \frac{1}{2}(k^{2}_{i}+k^{2}_{j})-\left[|E|^{-1}\sum\limits_{e_{ij}\in E} \frac{1}{2}(k_{i}+k_{j})\right]^{2}}
\end{equation}
in which $k_{i}$ is the degree of vertex $i$ and $e_{ij}$ denotes an edge connecting vertex $i$ to $j$.

By estimating the value for $r$ of graph $G(V,E)$, one can plausibly classify graph $G(V,E)$ into reasonable graph family. Graph $G(V,E)$ is considered assortative if $r$ is larger than zero, disassortative if $r$ is less than zero, non-assortative otherwise (i.e., $r=0$). In the same article \cite{Newman-2002}, Newman had experimentally shown that most social networks belong to the assortative mixing family and however almost all both biological and technological networks fall into the scope of disassortative mixing. In what follows, the task to answer is to determine which type of graph family the proposed graph $G^{\star}(m,t)$ is in.

\textbf{Proposition 4} The Pearson correlation coefficient $r(m;t)$ of graph $G^{\star}(m,t)$ can be written as

\begin{equation}\label{eqa:MF-3-4-2}
r(m;t)=\frac{\frac{\Lambda_{1}}{|E^{\star}(m,t)|}-\left(\frac{\Lambda_{2}}{|E^{\star}(m,t)|}\right)^{2}}{\frac{\Lambda_{3}}{|E^{\star}(m,t)|}-\left(\frac{\Lambda_{2}}{|E^{\star}(m,t)|}\right)^{2}},
\end{equation}
where $\Lambda_{1}=\sum\limits_{e_{ij}\in E^{\star}(m,t)} k_{i}k_{j}$, $\Lambda_{2}=\sum\limits_{e_{ij}\in E^{\star}(m,t)} \frac{1}{2}(k_{i}+k_{j})$, $\Lambda_{3}=\sum\limits_{e_{ij}\in E^{\star}(m,t)} \frac{1}{2}(k^{2}_{i}+k^{2}_{j})$, as well as  $$\Lambda_{1}=\left(\frac{m^{t+2}-1}{m-1}-1\right)\left[(2t+1)m^{t+1}+\frac{m^{t+1}-m}{m-1}\right]+m^{t+1}(t+1)\left(t+\frac{m^{t+1}-m}{m-1}\right),$$ $$\Lambda_{2}=\frac{m^{t+2}-1}{2(m-1)}\left(\frac{m^{t+1}-m}{m-1}+m^{t+1}\right)+\frac{m^{t+1}}{2}\left(t^{2}+4t+\frac{m^{t+1}-m}{m-1}\right),$$
$$\Lambda_{3}=\frac{1}{2}\left(\frac{m^{t+1}-m}{m-1}+m^{t+1}\right)\left(\frac{m^{t+2}-1}{m-1}-1\right)^{2}+\frac{(t+1)^{3}m^{t+1}}{2}+\frac{m^{t+1}}{2}\sum_{i=1}^{t}\frac{(m^{i}+1)^{3}}{m^{i}}.$$

\textbf{\emph{ Proof}} As explained above, it is sufficient to divide all the edges in graph $G^{\star}(m,t)$ into different classes according to two endpoint degrees of each edge. Again, we recall the concrete generation of graph $G^{\star}(m,t)$, which allows us to find that there are in fact number $2t+1$ of distinct edge subsets. Therefore, we will deal with such a computation for precisely deriving formula of Pearson correlation coefficient $r(m;t)$ as follows

\begin{itemize}
\item For the hub, there exist number $t+1$ of different types of edges connecting all other vertices but it in graph $G^{\star}(m,t)$ with respect to two endpoint degrees of edge. Let $E_{1}$ denote a set consisting of those edges. Accordingly, the number of edges whose another endpoint degree is equal to $m^{i}$ ($1\leq i\leq t$) and $t+1$ equal $m^{t+1-i}$ and $m^{t+1}$, respectively. Thus, one can have

  \begin{subequations}
\label{eq:whole}
\begin{eqnarray}
\Lambda_{11}=\sum\limits_{e_{ij}\in E_{1}} k_{i}k_{j}=\left(\frac{m^{t+2}-1}{m-1}-1\right)\left[\sum_{i=1}^{t}(m^{i}+1)m^{t+1-i}+m^{t+1}(t+1)\right],\label{subeq:MF-3-4-3}
\end{eqnarray}
\begin{equation}
\Lambda_{21}=\sum\limits_{e_{ij}\in E_{1}} \frac{1}{2}(k_{i}+k_{j})=\sum_{i=1}^{t}m^{t+1-i} \left(\frac{\frac{m^{t+2}-1}{m-1}+m^{i}}{2}\right)+m^{t+1}\left(\frac{t+\frac{m^{t+2}-1}{m-1}}{2}\right),\label{subeq:MF-3-4-4}
\end{equation}
\begin{equation}
\begin{aligned}\Lambda_{31}=\sum\limits_{e_{ij}\in E_{1}} \frac{1}{2}(k^{2}_{i}+k^{2}_{j})&=\left(\frac{m^{t+2}-1}{m-1}-1\right)^{2}\sum_{i=0}^{t}\frac{m^{i}}{2} +\frac{m^{t+1}(t+1)^{2}}{2}\\
&+\sum_{i=1}^{t}\frac{m^{t+1-i}(m^{i}+1)^{2}}{2}.\end{aligned}\label{subeq:MF-3-4-5}
\end{equation}
\end{subequations}

\item For an arbitrary vertex at the layer $i$ ($1\leq i\leq t$), there exist two distinct kinds of edges. The only one of those such edges has the other endpoint degree $\frac{m^{t+2}-1}{m-1}-1$. The contribution from such an edge to calculation of Pearson correlation coefficient $r(m;t)$ has been considered in the first case. Each of all other edges composing set $E_{2}$ connects to some vertex allocated at the layer $0$ and thus has the other endpoint degree equal to $t+1$. After that, one may obtain a group of equations

  \begin{subequations}
\label{eq:whole}
\begin{eqnarray}
\Lambda_{12}=\sum\limits_{e_{ij}\in E_{2}} k_{i}k_{j}=m^{t+1}\sum_{i=1}^{t}(t+1)(m^{i}+1),\label{subeq:MF-3-4-6}
\end{eqnarray}
\begin{equation}
\Lambda_{22}=\sum\limits_{e_{ij}\in E_{2}} \frac{1}{2}(k_{i}+k_{j})=m^{t+1}\sum_{i=1}^{t}\frac{(t+1)+(m^{i}+1)}{2},\label{subeq:MF-3-4-7}
\end{equation}
\begin{equation}
\Lambda_{32}=\sum\limits_{e_{ij}\in E_{2}} \frac{1}{2}(k^{2}_{i}+k^{2}_{j})=m^{t+1}\sum_{i=1}^{t}\frac{(t+1)^{2}+(m^{i}+1)^{2}}{2}.\label{subeq:MF-3-4-8}
\end{equation}
\end{subequations}

\end{itemize}

Till now, all possible cases are exhaustively enumerated. Next, we may obtain $\Lambda_{1}$ based on $\Lambda_{11}$ plus $\Lambda_{12}$ by using some simple arithmetics, namely, $\Lambda_{1}=\Lambda_{11}+\Lambda_{12}$. With an in spirit similar method to the above, the closed forms for $\Lambda_{2}$ and $\Lambda_{3}$ can be precisely calculated. Token together, we accomplish the proof of Proposition 4.

To better observe the behavior of Pearson correlation coefficient $r(m;t)$ in the limit of large graph size, we take as input the result from Eq.(\ref{eqa:MF-3-4-2}) and find a consequence shown in Fig.2. Apparently, in the limit of large graph size, the behavior of Pearson correlation coefficient $r(m;t)$ not only has a tendency to zero but also is always smaller than $0$. This reveals that our graph $G^{\star}(m,t)$ is disassortative.

\begin{figure}
\centering
  \includegraphics[height=7cm]{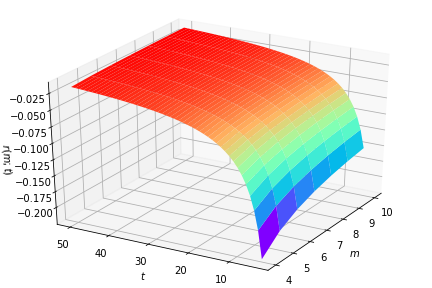}\\
{\small Fig.2. The diagram of Pearson correlation coefficient $r(m;t)$ of $G^{\star}(m,t)$. }
\end{figure}

\section{Average trapping time ($ATT$)}

Our main goal in the preceding section is to study some structural properties planted on the graph $G^{\star}(m,t)$. The results show that graph $G^{\star}(m,t)$ has some particular structural features, for instance, obeying power-law distribution with exponent $2$, unfound in most early models. In order to further probe many other properties of graph $G^{\star}(m,t)$, this section will aim at considering a special type of discrete-time unbiased random walk, i.e., the trapping problem, on graph $G^{\star}(m,t)$.

\subsection{Trapping problem}

The trapping problem on graph $G(V,E)$ is a special random walk with a deep trap allocated at an arbitrary vertex $\theta$($\in V$). Here, random walk is a paradigmatic dynamical process on graph $G(V,E)$ and has attracted a wide range of attentions in the past decades. It has been proven useful in many practical applications, such as, finding community structure on complex networks \cite{Rosvall-2008}. Mathematically, during a random walk on graph $G(V,E)$, a walker will hop on each vertex $v$ in its neighbor set with the probability $1/k_{u}$ from its current position $u$. Such a description can be interpreted by the transition matrix $\mathbf{P}=\mathbf{D}^{-1}\mathbf{A}$ of graph $G(V,E)$ where $\mathbf{A}$ is adjacency matrix defined as

$$a_{ij}=\left\{\begin{aligned}&1, \quad\text{vertex $i$ is adjacent to $j$}\\
&0,\quad\text{otherwise}
\end{aligned}\right.
$$
and $\mathbf{D}$ is the diagonal matrix which is referred to as $\mathbf{D}=\text{diag}[k_{1},k_{2},\dots,k_{|V|}]$ where the $i$th diagonal entry is $k_{i}=\sum_{j=1}^{|V|}a_{ij}$, while all non-diagonal elements are zero. After that, the jumping probability $P_{u\rightarrow v}$ for a walker starting out from $u$ to $v$ can be given by the following master equation

\begin{equation}\label{eqa:MF-4-1-0}
P_{u\rightarrow v}(t+1)=\sum_{i\in V}\frac{a_{iv}}{k_{i}}P_{u\rightarrow i}(t).
\end{equation}
In addition, there are more information about random walk on graph encapsulated in the transition matrix $\mathbf{P}$ above. For example, with the help of some techniques from matrix theory, one can study some quantities correlated to random walk of graph by matrix $\mathbf{P}$ more conveniently.

In this paper, we focus on a quantity associated with the trapping problem that is usually called the trapping time, denoted by $TT$. The $TT_{i\theta}$ is the expected time taken by a walker starting out from its present location $i$ to first reach to the trap $\theta$. As known, by using some methods from spectral graph theory \cite{Chung-1997}, the quantity $TT_{i\theta}$ can be written as

\begin{equation}\label{eqa:MF-4-1-1}
TT_{i\theta}=2|E|\sum_{s=2}^{|V|}\frac{1}{1-\lambda_{s}}\left(\frac{\psi_{si}}{k_{\theta}}-\frac{\psi_{s\theta}\psi_{si}}{\sqrt{k_{\theta}k_{i}}}\right),
\end{equation}
where we make use of some notations as follows, $1=\lambda_{1}>\lambda_{2}\geq \lambda_{3}\geq \dots\geq \lambda_{|V|}$ are the $|V|$ eigenvalues of matrix
$\mathbf{\Gamma}=\mathbf{D}^{\frac{1}{2}}\mathbf{P}\mathbf{D}^{-\frac{1}{2}}=\mathbf{D}^{-\frac{1}{2}}\mathbf{A}\mathbf{D}^{\frac{1}{2}}$ and $\psi_{1}, \psi_{2}, \psi_{3},\dots, \psi_{|V|}$ represent the corresponding mutually orthogonal eigenvectors of unit length in which $\psi_{i}=(\psi_{i1},\psi_{i2},\dots,\psi_{i|V|})$. For graph $G(V,E)$ as a whole, the average trapping time may be immediately obtained
\begin{equation}\label{eqa:MF-4-1-2}
ATT_{\theta}=\frac{1}{|V|-1}\sum_{i\in V, i\neq \theta} TT_{i\theta}.
\end{equation}

Inserting Eq.(\ref{eqa:MF-4-1-1}) into Eq.(\ref{eqa:MF-4-1-2}), the $ATT_{\theta}$ can be rearranged as

\begin{equation}\label{eqa:MF-4-1-3}
ATT_{\theta}=\frac{1}{1-\pi_{\theta}}\sum_{i=2}^{|V|}\left(\frac{1}{1-\lambda_{i}}\psi_{i\theta}^{2}\sum_{j=1}^{|V|}\frac{k_{j}}{k_{\theta}}\right)
\end{equation}
where $\pi_{\theta}$ is equal to $k_{\theta}/\sum_{j=1}^{|V|}k_{j}$. As said in \cite{Aldous-1999}, the theoretical lower bound for $ATT_{\theta}$ in Eq.(\ref{eqa:MF-4-1-3}) may be induced as

\begin{equation}\label{eqa:MF-4-1-4}
ATT_{\theta}\geq\frac{1}{1-\pi_{\theta}}\frac{\sum_{j=1}^{|V|}k_{j}}{k_{\theta}}(1-\pi_{\theta})^{2}=\frac{\sum_{j=1}^{|V|}k_{j}}{k_{\theta}}-1.
\end{equation}
As will be shown later, this is a sharp bound in the sense that it can be achieved in some graphs.

It is obvious to see that one of most important approaches to the task, calculating the average trapping time $ATT_{\theta}$ on a graph $G(V,E)$ with a trap $\theta$, is to first determine the adjacency matrix $\mathbf{A}$. In the past, it is using such a technique that the exact solutions to average trapping time on some graphs \cite{Zhang-2009,Benichoua-2014}, most of which are generated in an iterative fashion, have been derived.  While the proposed graph $G^{\star}(m,t)$ is also built up in an iterative manner like some previous graphs mentioned above, it is in practice slightly difficult to obtain the desirable solution by some arithmetics based on the adjacency matrix corresponding to graph $G^{\star}(m,t)$. To deal with this issue, we will take advantage of another tool based on master equation as in Eq.(\ref{eqa:MF-4-1-0}).

As a warm-up exercise before beginning with discussion on the trapping problem over scale-free graph $G^{\star}(m,t)$, let us consider some simple yet helpful examples. By using these examples, we will develop the first theorem in this paper.

\subsection{Star-type graph}

\emph{Example 1} For a complete graph $K_{n}$, the average trapping time $ATT_{1}$ to an arbitrary vertex $v$ selected as the trap $\theta$ is

\begin{equation}\label{eqa:MF-4-2-1-1}
ATT_{1}=n-1.
\end{equation}

\textbf{\emph{ Proof}} Here, let $x$ denote the trapping time for a walker starting out from a vertex $u$ ($\neq v$) to first hit the trap $\theta$. In terms of the structure of complete graph $K_{n}$, $x$ should follow

\begin{equation}\label{eqa:MF-4-2-1-2}
x=\frac{1}{n-1}+ \frac{n-2}{n-1}(1+x).
\end{equation}

Plugging the value of $x$ derived from Eq.(\ref{eqa:MF-4-2-1-2}) into Eq.(\ref{eqa:MF-4-1-2}) leads to the same result as that shown in Eq.(\ref{eqa:MF-4-2-1-1}). This is complete.

\emph{Example 2}  For a wheel graph $W_{(1,n)}$ \cite{Bondy-2008}, the average trapping time $ATT_{2}$ to the center $o$ chosen as the trap $\theta$ follows

\begin{equation}\label{eqa:MF-4-2-2-1}
ATT_{2}=3.
\end{equation}

\textbf{\emph{ Proof}} As above, let $y$ represent the trapping time for a walker starting out from a vertex $u$ ($\neq o$) to first reach to the trap $\theta$. Due to the structure of wheel graph $W_{(1,n)}$, $y$ obeys

\begin{equation}\label{eqa:MF-4-2-2-2}
y=\frac{1}{3}+ \frac{2}{3}(1+y).
\end{equation}
It is straightforward to obtain $y=3$. And then, by definition in Eq.(\ref{eqa:MF-4-1-2}), we may prove Eq.(\ref{eqa:MF-4-2-2-1}).

\emph{Example 3} For a star graph $S_{(1,n)}$, the average trapping time $ATT_{3}$ to the center $s$ designated as the trap $\theta$ satisfies

\begin{equation}\label{eqa:MF-4-2-3-1}
ATT_{3}=1.
\end{equation}

The proof of Eq.(\ref{eqa:MF-4-2-3-1}) can be easily obtained in a similar manner adopted in the two examples above and thus we omit detailed calculations.

Based on the demonstration in Eq.(\ref{eqa:MF-4-1-4}), it is not hard to show that the three graphs (i.e., $K_{n}$, $W_{(1,n)}$ and $S_{(1,n)}$) all have most optimal topological structure by achieving the theoretical lower bound for average trapping time in the trapping problem considered here. In fact, we can formalize the consequences obtained above to hold in a more general situation. One approach to this generalization is to seek for common attributes among the three graphs according to their own underlying structures. There have to be not much effort to see that they indeed share a structural feature with each other in common as will be shown shortly. In the following, we will introduce a formalized version of graphs of such kind.

\textbf{Definition} Given a regular graph $\mathcal{G}_{(n,m)}$ consisting of number $n$ of vertices with degree $m$ each, we may take an external vertex $u$ and then connect it to each vertex $v$ in graph $\mathcal{G}_{(n,m)}$ by one new edge, resulting in a new graph $\mathcal{G'}_{(n,m)}$ that is referred to as the \emph{star-type graph} in this paper.

By definition, complete graph $K_{n-1}$, cycle $C_{n}$ and empty graph $O_{n}$ are regular and so the three graphs mentioned in examples 1-3 all belong to star-type graph family. Fig.3 illustrates some concrete example graphs. In what follows, our goal is to generalize the aforementioned results to hold in star-type graph family $\mathcal{G'}_{(n,m)}$.

\begin{figure}
\centering
  \includegraphics[height=7cm]{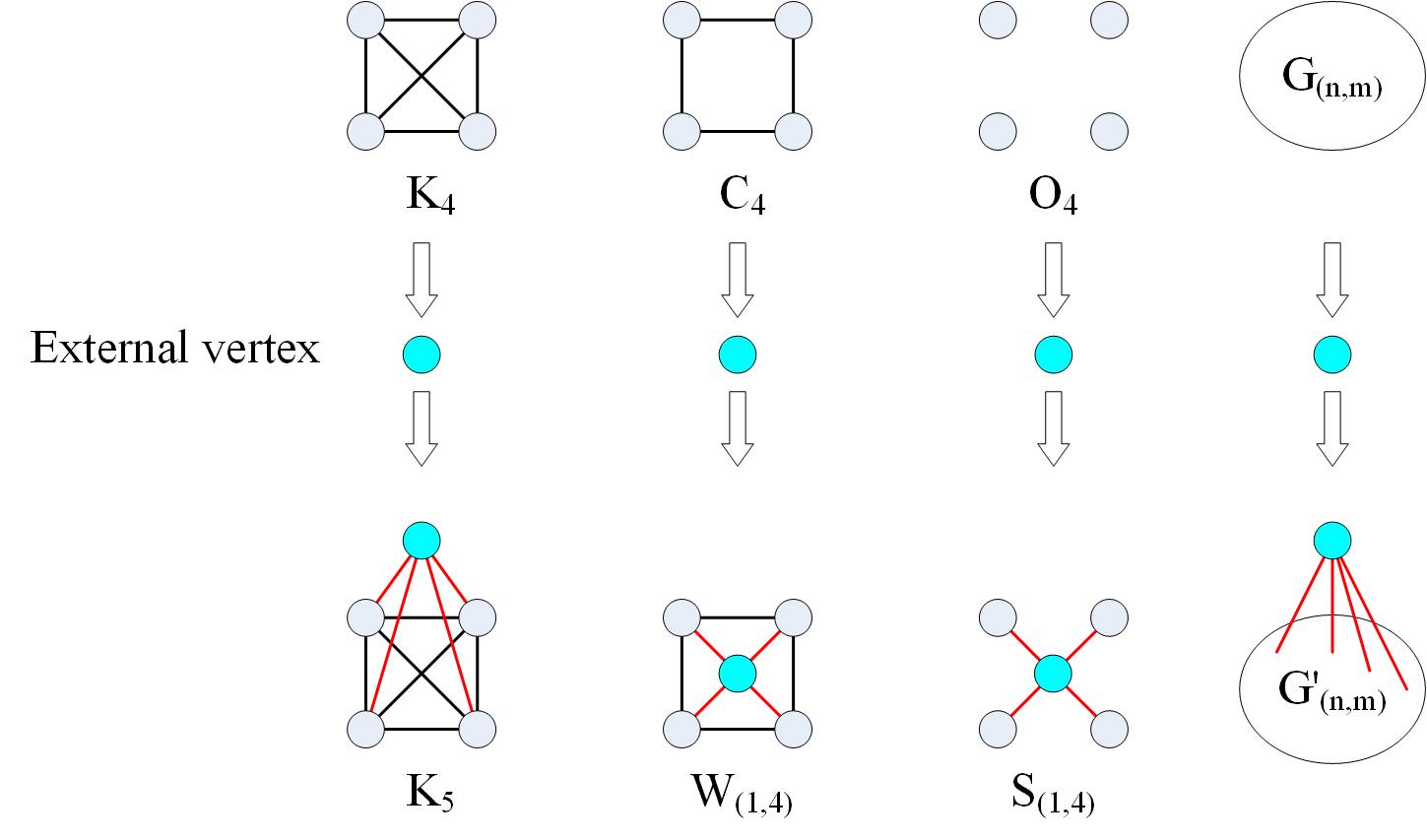}\\
{\small Fig.3. The diagram of several examples in star-type graph family $\mathcal{G'}_{(n,m)}$.  }
\end{figure}
\textbf{Theorem 1} For an arbitrary star-type graph $\mathcal{G'}_{(n,m)}$ upon regular graph $\mathcal{G}_{(n,m)}$, the solution to average trapping time $ATT$ to the center $u$, i.e., that newly added vertex, set to the trap $\theta$ is

\begin{equation}\label{eqa:MF-4-2-4-1}
ATT=m+1.
\end{equation}

\textbf{\emph{ Proof}} As previously, the trapping time $TT_{v}$ for each vertex $v$ but center $u$ can be written in the following term

\begin{equation}\label{eqa:MF-4-2-4-2}
TT_{v}=\frac{1}{m+1}+\frac{m}{m+1}(TT_{v}+1),
\end{equation}
here we already make use of the symmetric structure of regular graph $\mathcal{G}_{(n,m)}$. Solving for $TT_{v}$ from Eq.(\ref{eqa:MF-4-2-4-2}) yields
\begin{equation}\label{eqa:MF-4-2-4-3}
TT_{v}=m+1.
\end{equation}

Substituting the value for $TT_{v}$ into Eq.(\ref{eqa:MF-4-1-2}) produces the same result as that in Eq.(\ref{eqa:MF-4-2-4-1}), implying that Theorem 1 holds true. At the same time, all the star-type graphs $\mathcal{G'}_{(n,m)}$ show most optimal topological structure with respect to the trapping problem with a single trap where $ATT_{v}$ in Eq.(\ref{eqa:MF-4-2-4-1}) is identical to the theoretical lower bound calculated from Eq.(\ref{eqa:MF-4-1-4}).

Obviously, star-type graph $\mathcal{G'}_{(n,m)}$ displays some specific structural features, such as its seed $\mathcal{G}_{(n,m)}$ with symmetric character. In addition, as shown above, all vertices but that trap have the same trapping time $m+1$. Mathematically, the important reason for this is that these such vertices are all in a completely common local environment. In another words, it is the significantly specific structure of graph $\mathcal{G'}_{(n,m)}$ that plays a key role in the deductions of average trapping time. Thus, we can ask whether there are many other graphs with distinct topological structures from star-type graph $\mathcal{G'}_{(n,m)}$ whose average trapping times are subject to the theoretical lower bound in the trapping problem considered here. Towards this end, we will consider the trapping problem on scale-free graph $G^{\star}(m,t)$ built in Section 2 in more detail and then, fortunately, find graph $G^{\star}(m,t)$ to approximately achieve the the theoretical lower bound for average trapping time in the large graph size limit. This suggests that scale-free graph $G^{\star}(m,t)$ has some features of great interest that have not been observed in most networked graphs obeying power-law distribution \cite{Benichoua-2014}.

In the discussions above, the trap is always placed at that vertex with greatest degree in star-type graph $\mathcal{G'}_{(n,m)}$ mainly because in general, the greatest degree vertex is considered to play a crucial role in the trapping problem. Indeed, the results obtained illustrate this viewpoint. In other words, the analytic value for average trapping time is precisely equal to the theoretical lower bound when placing the trap at the greatest degree vertex in graph $\mathcal{G'}_{(n,m)}$. Motivated by this, in the subsequent sections, we will in-detail consider the trapping problem on scale-free graph $G^{\star}(m,t)$ in which some vertices with specific features, such as, greatest degree, may be chosen as candidates that are assigned traps.

\subsection{$ATT$ to hub vertex $\mathcal{H}(t)$}

From now on, let us turn our attention to discussions about the trapping problem on scale-free graph $G^{\star}(m,t)$.
Similar to the generation of star-type graph $\mathcal{G'}_{(n,m)}$, our scale-free graph $G^{\star}(m,t)$ is also constructed by connecting an external vertex to each vertex belonging to ingredients $G_{i}(m,t-1)$. Yet, there still exist some considerable differences between the both types of graphs. The most remarkable one compared to graph $\mathcal{G'}_{(n,m)}$ is the proposed graph $G^{\star}(m,t)$ has heterogeneous structure, i.e., a small fraction of vertices having a large number of connections. Specifically speaking, graph $G^{\star}(m,t)$ obeys power-law degree distribution, a feature that is popular in a great number of complex networks. As will be shown later, most interestingly, such a considerably structural property seems to have no prominent influence on the closed-form solution for average trapping time.

First of all, using Eq.(\ref{eqa:MF-4-1-4}), we can without difficulty obtain the theoretical lower bound for average trapping time $ATT_{\mathcal{H}(t)}$ when allocating a trap at the hub $\mathcal{H}(t)$, as below

\begin{equation}\label{eqa:MF-4-3-0}
ATT'_{\mathcal{H}(t)}=\frac{2|\mathcal{E}^{\star}_{m}(t)|}{k_{\mathcal{H}(t)}}-1\approx 2t-\frac{t-1}{m}+1.
\end{equation}

In fact, the approximate value for $ATT_{\mathcal{H}(t)}$ in the above equation can be reached with respect to analytical calculation to average trapping time. Below provides a detailed proof.

\textbf{Theorem 2} The closed-form solution to average trapping time $ATT_{\mathcal{H}(t)}$ is given by

\begin{equation}\label{eqa:MF-4-3-1}
ATT_{\mathcal{H}(t)}=O\left(t+\frac{1}{2}+\frac{1}{m}\right).
\end{equation}
In the limit of large $t$, the graph $G^{\star}(m,t)$ exhibits the approximately most optimal topological structure by means of holding the theoretical lower bound for $ATT_{\mathcal{H}(t)}$ in Eq.(\ref{eqa:MF-4-3-0}).

\textbf{\emph{ Proof}} Before proceeding further, we need to come up with some notations that help to accomplish the proof to Eq.(\ref{eqa:MF-4-3-1}). Let symbol $P_{\mathcal{H}(t)}(L;s)$ $(1\leq L \leq t+1)$ denote the probability for a walker originally set to a vertex $u$ at the layer $L$ to first visit that trap $\mathcal{H}(t)$ in $s$ steps. And then, it is clear to the eye that probabilities $P_{h}(L;s)$ $(1\leq L \leq t+1)$ satisfy the following master equations

\begin{subequations}
\label{eq:whole}
\begin{eqnarray}
P_{\mathcal{H}(t)}(t+1;s)=\frac{\delta_{s,1}}{k_{t+1}}+\frac{1}{k_{t+1}}\sum_{L=1}^{t}P_{\mathcal{H}(t)}(L;s-1),\label{subeq:MF-4-3-2}
\end{eqnarray}
\begin{equation}
P_{\mathcal{H}(t)}(L;s)=\frac{\delta_{s,1}}{k_{L}}+\frac{k_{L}-1}{k_{L}}P_{\mathcal{H}(t)}(t+1;s-1).\label{subeq:MF-4-3-3}
\end{equation}
\end{subequations}
Here $k_{L}$ is the degree of vertex lying at the layer $L$ $(1\leq L \leq t+1)$ and $\delta_{s,1}$ represents the Kronecker delta function where $\delta_{s,1}$ is equal to $1$ when $s=1$ and $0$ otherwise. In general, one well-used approach to addressing the both equations above is probability generating function. Therefore, for our purpose, we denote by $\mathcal{P}_{\mathcal{H}(t)}(L;z)$ the probability generating function corresponding to probability $P_{\mathcal{H}(t)}(L;s)$. After that, using methods from probability generating function, Eqs.(\ref{subeq:MF-4-3-2}) and (\ref{subeq:MF-4-3-3}) are able to be rearranged as

\begin{subequations}
\label{eq:whole}
\begin{eqnarray}
\mathcal{P}_{\mathcal{H}(t)}(t+1;z)=\frac{z}{k_{t+1}}+\frac{z}{k_{t+1}}\sum_{L=1}^{t}\mathcal{P}_{\mathcal{H}(t)}(L;z),\label{subeq:MF-4-3-4}
\end{eqnarray}
\begin{equation}
\mathcal{P}_{\mathcal{H}(t)}(L;z)=\frac{z}{k_{L}}+\frac{(k_{L}-1)z}{k_{L}}\mathcal{P}_{\mathcal{H}(t)}(t+1;z).\label{subeq:MF-4-3-5}
\end{equation}
\end{subequations}

Using a trivial fact associated to probability generating function $\mathcal{P}_{\mathcal{H}(t)}(t+1;z)$, that is, the first-order differential of $\mathcal{P}_{\mathcal{H}(t)}(t+1;z)$ at value $z=1$ defined to be the expected trapping time as follows

\begin{equation}\label{eqa:MF-4-3-6}
TT_{\mathcal{H}(t)}(t+1)=\left.\dfrac{d}{dz}\mathcal{P}_{\mathcal{H}(t)}(t+1;z)\right|_{z=1},
\end{equation}
we can perform differential on the both sides of Eqs.(\ref{subeq:MF-4-3-4}) and (\ref{subeq:MF-4-3-5}) and then write a series of iterative equations

\begin{subequations}
\label{eq:whole}
\begin{eqnarray}
TT_{\mathcal{H}(t)}(t+1)=\left.\dfrac{d}{dz}\mathcal{P}_{h}(t+1;z)\right|_{z=1}=1+\frac{1}{k_{t+1}}\sum_{L=1}^{t}TT_{\mathcal{H}(t)}(L),
\label{subeq:MF-4-3-7}
\end{eqnarray}
\begin{equation}
TT_{\mathcal{H}(t)}(L)=\left.\dfrac{d}{dz}\mathcal{P}_{h}(L;z)\right|_{z=1}=1+\frac{k_{L}-1}{k_{L}}TT_{\mathcal{H}(t)}(t+1).\label{subeq:MF-4-3-8}
\end{equation}
\end{subequations}

According to Eqs.(\ref{subeq:MF-4-3-7}) and (\ref{subeq:MF-4-3-8}), the exact solution for trapping time $TT_{\mathcal{H}(t)}(t+1)$ can be solved to express

\begin{equation}\label{eqa:MF-4-3-9}
TT_{\mathcal{H}(t)}(t+1)=\frac{2t+1}{t+1-\sum_{L=1}^{t}\frac{k_{L}-1}{k_{L}}}.
\end{equation}

Thus,

\begin{equation}\label{eqa:MF-4-3-10}
TT_{\mathcal{H}(t)}(L)=1+\frac{k_{L}-1}{k_{L}}\left(\frac{2t+1}{t+1-\sum_{L=1}^{t}\frac{k_{L}-1}{k_{L}}}\right).
\end{equation}

Substituting Eqs.(\ref{eqa:MF-4-3-9}) and (\ref{eqa:MF-4-3-10}) into Eq.(\ref{eqa:MF-4-1-2}) produces the precise solution for $ATT_{\mathcal{H}(t)}$

\begin{equation}\label{eqa:MF-4-3-11}
TT_{\mathcal{H}(t)}(L)\approx t+\frac{1}{2}+\frac{1}{m}.
\end{equation}
This proves Theorem 2.

By far, the trapping problem with a single trap has been completely answered. Additionally, we are also interested in the multi-traps problem since many real-world applications, for example, target search, contain many destination objects that might be abstractly thought of as the traps considered in this paper. Hence, to some extents, it is quite meaningful to handle the issue with multi-traps. The following subsections aim at reporting some consequences correlated to the multi-traps problem on the proposed graph $G^{\star}(m,t)$.

\subsection{$ATT$ to multiple vertices}

As said above, one of most significant reasons for assigning the trap at vertex $\mathcal{H}(t)$ is that all vertices but vertex $\mathcal{H}(t)$ in graph $G^{\star}(m,t)$ are connected to the hub. It is obvious to see that an arbitrary vertex $v$ at the layer $L$ ($0\leq L\leq t$) is also connected to some vertex at the bottom-most layer. If we unify all the vertices at the layer $t+1$ into a hyper-vertex $\alpha$, then one finds that the newly unified vertex  $\alpha$ connects to all other vertices in graph $G^{\star}(m,t)$. Similarly, we want to learn about what the average trapping time $ATT_{\alpha}$ is when the hyper-vertex $\alpha$ is occupied by number $m^{t+1}$ of traps.

\subsubsection{$ATT$ to hyper-vertex $\alpha$}

With the help of Eq.(\ref{eqa:MF-4-1-4}), in such a situation where each vertex at the bottom-most layer is filled with one trap, we can calculate the the theoretical lower bound for average trapping time $ATT_{\alpha}$ and write

\begin{equation}\label{eqa:MF-4-4-1-0}
ATT'_{\alpha}=\frac{2|\mathcal{E}^{\star}_{m}(t)|}{m^{t+1}(t+1)}-1=O\left(1+\frac{1}{t}\right).
\end{equation}

And then, the analytic value for $ATT_{\alpha}$ is follows.

\textbf{Theorem 3} The closed-form solution to average trapping time $ATT_{\alpha}$ is calculated equal to

\begin{equation}\label{eqa:MF-4-4-1-1}
ATT_{\alpha}=\frac{\frac{1+\sum_{i=1}^{t}\frac{m^{i}}{k_{0}}}{1-\frac{1}{k_{0}}\sum_{i=1}^{t}\frac{m^{i}}{k_{i}}}+\sum_{i=1}^{t}m^{i}\left[1+\frac{1}{k_{i}}\left(\frac{1+\sum_{i=1}^{t}\frac{m^{i}}{k_{0}}}{1-\frac{1}{k_{0}}\sum_{i=1}^{t}\frac{m^{i}}{k_{i}}}\right)\right]}{\frac{m^{t+2}-1}{m-1}-m^{t+1}},
\end{equation}
where $k_{i}$ ($0\leq i\leq t$) is the degree of vertex at the layer $i$.

\textbf{\emph{ Proof}} As previously, we have to introduce some notations as follows (i) symbol $P_{\alpha}(0;s)$ represents the probability that a walker at the hub first hop on some vertex at the layer $t+1$ in $s$ steps and (ii) $P_{\alpha}(L;s)$ is viewed as the probability for a walker at an arbitrary vertex at the layer $L$ ($1\leq L\leq t$) to first hit some vertex at the bottom-most layer. And then, we can obtain a group of master equations

\begin{subequations}
\label{eq:whole}
\begin{eqnarray}
P_{\alpha}(0;s)=\frac{m^{t+1}}{k_{0}}\delta_{s,1}+\sum_{i=1}^{t}\frac{m^{i}}{k_{0}}P_{\alpha}(i;s-1),\label{subeq:MF-4-4-1-2}
\end{eqnarray}
\begin{equation}
P_{\alpha}(i;s)=\frac{k_{i}-1}{k_{i}}\delta_{s,1}+\frac{1}{k_{i}}P_{\alpha}(0;s-1),\label{subeq:MF-4-4-1-3}
\end{equation}
\end{subequations}
in which $\delta_{s,1}$ is the Kronecker delta function as above and $k_{i}$ is the degree of vertex at the layer $i$.

Using technique from probability generating function, we denote by $\mathcal{P}_{\alpha}(i;z)$ generating function associated with the probability $P_{\alpha}(i;s)$. After that, Eqs.(\ref{subeq:MF-4-4-1-2}) and (\ref{subeq:MF-4-4-1-3}) can be rewritten as

\begin{subequations}
\label{eq:whole}
\begin{eqnarray}
\mathcal{P}_{\alpha}(0;z)=\frac{m^{t+1}}{k_{0}}z+\sum_{i=1}^{t}\frac{m^{i}z}{k_{0}}\mathcal{P}_{\alpha}(i;s),\label{subeq:MF-4-4-1-4}
\end{eqnarray}
\begin{equation}
\mathcal{P}_{\alpha}(i;z)=\frac{k_{i}-1}{k_{i}}z+\frac{z}{k_{i}}\mathcal{P}_{\alpha}(0;s).\label{subeq:MF-4-4-1-5}
\end{equation}
\end{subequations}

Similar to the development of Eq.(\ref{eqa:MF-4-3-6}), taking differential on the both sides of Eqs.(\ref{subeq:MF-4-4-1-4}) and (\ref{subeq:MF-4-4-1-5}) produces, respectively,

\begin{subequations}
\label{eq:whole}
\begin{eqnarray}
TT_{\alpha}(0)=\left.\dfrac{d}{dz}\mathcal{P}_{\alpha}(0;z)\right|_{z=1}=1+\sum_{i=1}^{t}\frac{m^{i}}{k_{0}}TT_{\alpha}(i),\label{subeq:MF-4-4-1-6}
\end{eqnarray}
\begin{equation}
TT_{\alpha}(i)=\left.\dfrac{d}{dz}\mathcal{P}_{\alpha}(i;z)\right|_{z=1}=1+\frac{1}{k_{i}}TT_{\alpha}(0)\label{subeq:MF-4-4-1-7}
\end{equation}
\end{subequations}
where $TT_{\alpha}(i)$ denotes the expected time taken by a walker originally starting out from one vertex allocated at the layer $i$ to first visit some vertex lying at the layer $t+1$.

Inserting Eq.(\ref{subeq:MF-4-4-1-7}) into Eq.(\ref{subeq:MF-4-4-1-6}) outputs

\begin{equation}\label{eqa:MF-4-4-1-8}
TT_{\alpha}(0)=\frac{k_{0}+\sum_{i=1}^{t}m^{i}}{k_{0}-\sum_{i=1}^{t}\frac{m^{i}}{k_{i}}}.
\end{equation}

And then, for $1\leq i\leq t$, we have

\begin{equation}\label{eqa:MF-4-4-1-9}
TT_{\alpha}(i)=1+\frac{1}{k_{i}}\left(\frac{k_{0}+\sum_{i=1}^{t}m^{i}}{k_{0}-\sum_{i=1}^{t}\frac{m^{i}}{k_{i}}}\right).
\end{equation}

Upon Eq.(\ref{eqa:MF-4-1-2}), the precise formula for average trapping time $ATT_{\alpha}$ can be obtained

\begin{equation}\label{eqa:MF-4-4-1-10}
ATT_{\alpha}=\frac{\frac{k_{0}+\sum_{i=1}^{t}m^{i}}{k_{0}-\sum_{i=1}^{t}\frac{m^{i}}{k_{i}}}+\sum_{i=1}^{t}m^{i}\left[1+\frac{1}{k_{i}}\left(\frac{k_{0}+\sum_{i=1}^{t}m^{i}}{k_{0}-\sum_{i=1}^{t}\frac{m^{i}}{k_{i}}}\right)\right]}{\frac{m^{t+2}-1}{m-1}-m^{t+1}},
\end{equation}
in which we have made use of a fact that the total number of vertices with trapping time $TT_{\alpha}(i)$ is equivalent to $m^{i}$.

Using some simple arithmetics, the result from Eq.(\ref{eqa:MF-4-4-1-10}) reduces to that of Eq.(\ref{eqa:MF-4-4-1-1}). This proves Theorem 3.

Nonetheless, we want to know about whether such an analytic value $ATT_{\alpha}$ in Eq.(\ref{eqa:MF-4-4-1-1}) is asymptotically equal to the theoretical lower bound $ATT'_{\alpha}$ in Eq.(\ref{eqa:MF-4-4-1-0}) in the limit of large $t$. To this end, we denote by parameter $\Delta_{\alpha}$ the ratio of $ATT_{\alpha}$ and $ATT'_{\alpha}$ and then feed the three parameters, i.e., ratio $\Delta_{\alpha}$, $m$ and $t$, into computer. As illustrated in Fig.4, our scale-free graph $G^{\star}(m,t)$ still exhibits the more optimal trapping efficiency when we assign the traps to all vertices at the bottom-most layer.

\begin{figure}
\centering
  \includegraphics[height=7cm]{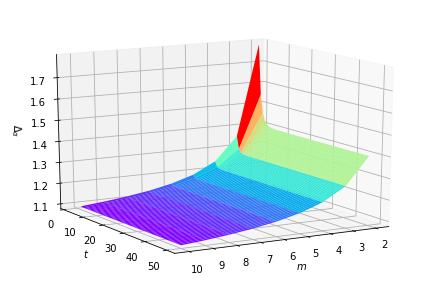}\\
{\small Fig.4. The diagram of ratio $\Delta_{\alpha}$. In the limit of large $t$, the value for $\Delta_{\alpha}$ tends to $1$, implying that the analytic solution to $ATT_{\alpha}$ is approximately close to the theoretical lower bound $ATT'_{\alpha}$.   }
\end{figure}

By the theoretical analysis above, the hub acts as a similar role to that hyper-vertex $\alpha$ since the average average trapping times corresponding to the both vertices hold on the theoretical lower bound, separately. Compered to the demonstration in Theorem 2, the result in Theorem 3 shows that choosing all vertices at the layer $t+1$ as potential trap locations leads to the most optimal trapping efficiency. In another words, on average, a walker from an arbitrary vertex at the layer $i$ ($0\leq i\leq t$) takes only one step to first hit some vertex at the layer $t+1$. Generally speaking, one hop should be the smallest measure for a walker that is doing random walk in the trapping problem. Hence, we are able to state that the assignment of traps to the hyper-vertex $\alpha$ is perfect$\footnote{In the trapping problem on a graph $G(V,E)$, a scheme $\xi$ for assigning traps to a set of vertices $V_{1}$ ($\neq V$) may be considered perfect such that a walker coming from an arbitrary vertex $v\in V-V_{1}$ spends only one step arriving some vertex $u\in V_{1}$ on average.}$ in the trapping problem considered in subsection 4.4.1.

As reported above, the potential trap locations have crucial effect on average trapping time. For a given graph $G(V,E)$, it is intriguing and challenging to seek for a perfect assignment of traps in the trapping problem. In fact, such an issue is closely related to one of the classic combinatorics problems, i.e., looking for maximum independent set on graph in question. As known, there is no polynomial algorithm for solving the problem of determining maximum independent set of a given graph. Here, based on both the specific structure of our graph $G^{\star}(m,t)$ and the concrete expression of average trapping time $ATT_{\alpha}$, we find the hyper-vertex $\alpha$ to be the maximum independent set of graph $G^{\star}(m,t)$. On the other hand, there are significant differences between the both problems in nature. The scheme for placing traps at a maximum independent set of a graph is not always perfect. For example, as discussed above, the maximum independent set of complete graph $K_{n}$ consists only of a single vertex and however Eq.(\ref{eqa:MF-4-2-1-1}) says that designating the trap at that vertex is not perfect. Yet, star graph $S_{(1,n)}$ in practice has a perfect assignment scheme that is not that considered in example 3. This perfect project, however, is to allocate traps at each leaf vertex that in fact together constitute the maximum independent set of star graph $S_{(1,n)}$ \cite{Definition-2020-1}. In a word, we would like to stress that finding perfect trap assignment has considerably practical applications and greatly theoretical flavor. As future direction, we will pay more attention to building up available approaches to addressing this interesting and challenging issue for more general graphs. At the same time, we also wish to witness some meaningful consequences related to perfect trap assignment in the trapping problem developed by other interested researchers in the days to come.

By far, we already study the trapping problem on the proposed graph $G^{\star}(m,t)$ where traps are placed at some vertices of great interest, such as with largest degree or together spanning a maximum independent set. Some related results have been listed above. In order to make progress further, the next subsection will continue to consider the trapping problem with multi-traps. By analogy with the discussions above, the trap is now distributed to each vertex belonging to some designated layer $i$ ($1\leq i\leq t$). Here, those vertices at the layer $i$ can be identified to a hyper-vertex $\beta_{i}$ ($1\leq i\leq t$) for convenience too.

\subsubsection{$ATT$ to hyper-vertex $\beta_{j}$}

Along the same line as the aforementioned subsection, the first task is to calculate the theoretical lower bound for average trapping time $ATT_{\beta_{j}}$ on the basis of Eq.(\ref{eqa:MF-4-1-4}), as below

 \begin{equation}\label{eqa:MF-4-4-2-0}
ATT'_{\beta_{j}}=\frac{2|\mathcal{E}^{\star}_{m}(t)|}{m^{t+1}+m^{j}}-1=O(2t+1).
\end{equation}
This has the same order of magnitude as that of Eq.(\ref{eqa:MF-4-3-0}) in the limit of large graph size.

Interestingly, the value for $ATT'_{\beta_{j}}$ can also be approximately hold by the analytic value for average trapping time $ATT_{\beta_{j}}$ in the limit of large $t$ as will be proved later. For our purpose, below puts forward some useful notations $\Omega_{i}$ ($i=0,1,\dots, 4$) and $\Psi_{i}$ ($i=0,1$) whose corresponding implications will be clarified in the process of validating Theorem 4.

 \begin{equation}\label{eqa:MF-4-4-2-1}
\left\{\begin{aligned}
&\Omega_{0}=\sum_{i=1}^{j-1}\frac{1}{k_{i}}+\sum_{i=j+1}^{t}\frac{1}{k_{i}},\qquad \Omega_{1}=\sum_{i=1}^{j-1}\frac{k_{i}-1}{k_{i}}+\sum_{i=j+1}^{t}\frac{k_{i}-1}{k_{i}},\\
&\Omega_{2}=\sum_{i=1}^{j-1}\frac{m^{i}}{k_{0}}+\sum_{i=j+1}^{t}\frac{m^{i}}{k_{0}},\qquad \Omega_{3}=\frac{1}{k_{0}}\left(\sum_{i=1}^{j-1}\frac{m^{i}}{k_{i}}+\sum_{i=j+1}^{t}\frac{m^{i}}{k_{i}}\right),\\
&\Omega_{4}=\sum_{i=1}^{j-1}\frac{m^{i}}{k_{0}}\frac{k_{i}-1}{k_{i}}+\sum_{i=j+1}^{t}\frac{m^{i}}{k_{0}}\frac{k_{i}-1}{k_{i}},\\
&\Psi_{1}=\frac{1+\Omega_{2}+\left(\frac{m^{t+1}}{k_{0}}+\Omega_{4}\right)\left(\frac{k_{t+1}+t-1}{k_{t+1}}\right)}{1-\left[\Omega_{3}+\left(\frac{m^{t+1}}{k_{0}}+\Omega_{4}\right)\left(\frac{1+\Omega_{0}}{k_{t+1}-\Omega_{1}}\right)\right]},\\
&\Psi_{2}=\frac{k_{t+1}+t-1}{k_{t+1}-\Omega_{1}}+\left(\frac{1+\Omega_{0}}{k_{t+1}-\Omega_{1}}\right)\Psi_{1},
\end{aligned}\right.
\end{equation}
here $k_{i}$ indicates the degree of vertex at the layer $i$ $(0\leq i\leq t+1)$ as above.

\textbf{Theorem 4} The closed-form solution to average trapping time $ATT_{\beta_{i}}$ is written in the next term

\begin{equation}\label{eqa:MF-4-4-2-2}
ATT_{\beta_{j}}=\frac{m^{t+1}-m^{j+1}+m^{j}-m+(m-1)[(m^{t+1}+k_{0}\Omega_{4})\Psi_{2}+(1+k_{0}\Omega_{3})\Psi_{1}]}{m^{t+2}-m^{j+1}+m^{j}-1}.
\end{equation}

\textbf{\emph{ Proof}} Using a similar approach to proving Theorem 3, let notations $P_{\beta_{j}}(t+1;s)$, $P_{\beta_{j}}(i;s)$ $(i\neq j)$ as well as $P_{\beta_{j}}(0;s)$, respectively, denote the probability that a walker moves from its original location $v$, a vertex that is at the layer $i$, to some trap in the layer $j$ in $s$ steps. And then, these probabilities satisfy the following master equations

\begin{subequations}
\label{eq:whole}
\begin{eqnarray}
P_{\beta_{j}}(t+1;s)=\frac{\delta_{s,1}}{k_{t+1}}+\left(\frac{1}{k_{t+1}}\sum_{i=0}^{j-1}P_{\beta_{j}}(i;s-1)+\frac{1}{k_{t+1}}\sum_{i=j+1}^{t}P_{\beta_{j}}(i;s-1)\right),\label{subeq:MF-4-4-2-3}
\end{eqnarray}
\begin{equation}
P_{\beta_{j}}(i;s)=\frac{1}{k_{i}}P_{\beta_{j}}(0;s-1)+\frac{k_{i}-1}{k_{i}}P_{\beta_{j}}(t+1;s-1),\label{subeq:MF-4-4-2-4}
\end{equation}
\begin{equation}
P_{\beta_{j}}(0;s)=\frac{m^{j}}{k_{0}}\delta_{s,1}+\left(\sum_{i=1}^{j-1}\frac{m^{i}}{k_{0}}P_{\beta_{j}}(i;s-1)+\sum_{i=j+1}^{t+1}\frac{m^{i}}{k_{0}}P_{\beta_{j}}(i;s-1)\right).\label{subeq:MF-4-4-2-5}
\end{equation}
\end{subequations}

Following Eqs.(\ref{subeq:MF-4-4-2-3})-(\ref{subeq:MF-4-4-2-5}), the probability generating functions corresponding to quantities $P_{\beta_{j}}(t+1;s)$, $P_{\beta_{j}}(i;s)$ and $P_{\beta_{j}}(0;s)$ can be obtained

\begin{subequations}
\label{eq:whole}
\begin{eqnarray}
\mathcal{P}_{\beta_{j}}(t+1;z)=\frac{z}{k_{t+1}}+\left(\frac{z}{k_{t+1}}\sum_{i=0}^{j-1}\mathcal{P}_{\beta_{j}}(i;z)+\frac{z}{k_{t+1}}\sum_{i=j+1}^{t}\mathcal{P}_{\beta_{j}}(i;z)\right),\label{subeq:MF-4-4-2-6}
\end{eqnarray}
\begin{equation}
\mathcal{P}_{\beta_{j}}(i;z)=\frac{z}{k_{i}}\mathcal{P}_{\beta_{j}}(0;z)+\frac{(k_{i}-1)z}{k_{i}}\mathcal{P}_{\beta_{j}}(t+1;z),\label{subeq:MF-4-4-2-7}
\end{equation}
\begin{equation}
\mathcal{P}_{\beta_{j}}(0;z)=\frac{m^{j}}{k_{0}}z+\left(\sum_{i=1}^{j-1}\frac{m^{i}z}{k_{0}}\mathcal{P}_{\beta_{j}}(i;z)+\sum_{i=j+1}^{t+1}\frac{m^{i}z}{k_{0}}\mathcal{P}_{\beta_{j}}(i;z)\right).\label{subeq:MF-4-4-2-8}
\end{equation}
\end{subequations}

After performing differential on the two sides of Eqs.(\ref{subeq:MF-4-4-2-6})-(\ref{subeq:MF-4-4-2-8}), the expected trapping time $TT_{\beta_{j}}(i)$ ($i\neq j$) can be derived

\begin{subequations}
\label{eq:whole}
\begin{eqnarray}
\begin{aligned}TT_{\beta_{j}}(t+1)&=\left.\dfrac{d}{dz}\mathcal{P}_{\beta_{j}}(t+1;z)\right|_{z=1}\\
&=1+\frac{1}{k_{t+1}}TT_{\beta_{j}}(0)+\frac{1}{k_{t+1}}\left(\sum_{i=1}^{j-1}TT_{\beta_{j}}(i)+\sum_{i=j+1}^{t}TT_{\beta_{j}}(i)\right)
\end{aligned},\label{subeq:MF-4-4-2-9}
\end{eqnarray}
\begin{equation}
TT_{\beta_{j}}(i)=\left.\dfrac{d}{dz}\mathcal{P}_{\beta_{j}}(i;z)\right|_{z=1}=1+\frac{1}{k_{i}}TT_{\beta_{j}}(0)+\frac{k_{i}-1}{k_{i}}TT_{\beta_{j}}(t+1),\label{subeq:MF-4-4-2-10}
\end{equation}
\begin{equation}
\begin{aligned}TT_{\beta_{j}}(0)&=\left.\dfrac{d}{dz}\mathcal{P}_{\beta_{j}}(0;z)\right|_{z=1}\\
&=1+\frac{m^{t+1}}{k_{0}}TT_{\beta_{j}}(t+1)+\left(\sum_{i=1}^{j-1}\frac{m^{i}}{k_{0}}TT_{\beta_{j}}(i)+\sum_{i=j+1}^{t}\frac{m^{i}}{k_{0}}TT_{\beta_{j}}(i)\right)
\end{aligned}.\label{subeq:MF-4-4-2-11}
\end{equation}
\end{subequations}

For brevity, we now introduce two notations $\Omega_{0}=\sum_{i=1}^{j-1}\frac{1}{k_{i}}+\sum_{i=j+1}^{t}\frac{1}{k_{i}}$ and $\Omega_{1}=\sum_{i=1}^{j-1}\frac{dk_{i}-1}{k_{i}}+\sum_{i=j+1}^{t}\frac{k_{i}-1}{k_{i}}$. After that, Eq.(\ref{subeq:MF-4-4-2-9}) can be reorganized as

\begin{equation}\label{eqa:MF-4-4-2-12}
TT_{\beta_{j}}(t+1)=1+\frac{t-1}{k_{t+1}}+\frac{1+\Omega_{0}}{k_{t+1}}TT_{\beta_{j}}(0)+\frac{\Omega_{1}}{k_{t+1}}TT_{\beta_{j}}(t+1).
\end{equation}

Analogously, according to three new notations, $\Omega_{2}=\sum_{i=1}^{j-1}\frac{m^{i}}{k_{0}}+\sum_{i=j+1}^{t}\frac{m^{i}}{k_{0}}$, $\Omega_{3}=\frac{1}{k_{0}}\left(\sum_{i=1}^{j-1}\frac{m^{i}}{k_{i}}+\sum_{i=j+1}^{t}\frac{m^{i}}{k_{i}}\right)$ along with $\Omega_{4}=\sum_{i=1}^{j-1}\frac{m^{i}}{k_{0}}\frac{k_{i}-1}{k_{i}}+\sum_{i=j+1}^{t}\frac{m^{i}}{k_{0}}\frac{k_{i}-1}{k_{i}}$, Eq.(\ref{subeq:MF-4-4-2-11}) may be converted into

\begin{equation}\label{eqa:MF-4-4-2-13}
TT_{\beta_{j}}(0)=1+\Omega_{2}+\left(\frac{m^{t+1}}{k_{0}}+\Omega_{4}\right)TT_{\beta_{j}}(t+1)+\Omega_{3}TT_{\beta_{j}}(0).
\end{equation}

In terms to Eqs.(\ref{eqa:MF-4-4-2-12}) and (\ref{eqa:MF-4-4-2-13}), we may use simple arithmetics to rearrange $TT_{\beta_{j}}(0)$ as

\begin{equation}\label{eqa:MF-4-4-2-14}
\begin{aligned}TT_{\beta_{j}}(0)&=1+\Omega_{2}+\left(\frac{m^{t+1}}{k_{0}}+\Omega_{4}\right)\left(\frac{k_{t+1}+t-1}{k_{t+1}}\right)\\
&+\left[\Omega_{3}+\left(\frac{m^{t+1}}{k_{0}}+\Omega_{4}\right)\left(\frac{1+\Omega_{0}}{k_{t+1}-\Omega_{1}}\right)\right]TT_{\beta_{j}}(0).
\end{aligned}\end{equation}

In what follows, the precise solution for trapping time $TT_{\beta_{j}}(0)$ is

\begin{equation}\label{eqa:MF-4-4-2-15}
TT_{\beta_{j}}(0)=\frac{1+\Omega_{2}+\left(\frac{m^{t+1}}{k_{0}}+\Omega_{4}\right)\left(\frac{k_{t+1}+t-1}{k_{t+1}}\right)}{1-\left[\Omega_{3}+\left(\frac{m^{t+1}}{k_{0}}+\Omega_{4}\right)\left(\frac{1+\Omega_{0}}{k_{t+1}-\Omega_{1}}\right)\right]}.
\end{equation}

Plugging the value for $TT_{\beta_{j}}(0)$ into Eq.(\ref{eqa:MF-4-4-2-12}) yields

\begin{equation}\label{eqa:MF-4-4-2-16}
TT_{\beta_{j}}(t+1)=\frac{k_{t+1}+t-1}{k_{t+1}-\Omega_{1}}+\left(\frac{1+\Omega_{0}}{k_{t+1}-\Omega_{1}}\right)\frac{1+\Omega_{2}+\left(\frac{m^{t+1}}{k_{0}}+\Omega_{4}\right)\left(\frac{k_{t+1}+t-1}{k_{t+1}}\right)}{1-\left[\Omega_{3}+\left(\frac{m^{t+1}}{k_{0}}+\Omega_{4}\right)\left(\frac{1+\Omega_{0}}{k_{t+1}-\Omega_{1}}\right)\right]}.
\end{equation}

Taking into account Eqs.(\ref{subeq:MF-4-4-2-10}), (\ref{eqa:MF-4-4-2-15}) and (\ref{eqa:MF-4-4-2-16}), by definition in Eq.(\ref{eqa:MF-4-1-2}), we finally derive the closed-form solution to average trapping time $ATT_{\beta_{j}}$

\begin{equation}\label{eqa:MF-4-4-2-17}
ATT_{\beta_{j}}=\frac{\left(\frac{m^{t+1}-m}{m-1}-m^{j}\right)+(m^{t+1}+k_{0}\Omega_{4})TT_{\beta_{j}}(t+1)+(1+k_{0}\Omega_{3})TT_{\beta_{j}}(0)}{\frac{m^{t+2}-1}{m-1}-m^{j}}.
\end{equation}

Utilizing two notations $\Psi_{1}=\frac{1+\Omega_{2}+\left(\frac{m^{t+1}}{k_{0}}+\Omega_{4}\right)\left(\frac{k_{t+1}+t-1}{k_{t+1}}\right)}{1-\left[\Omega_{3}+\left(\frac{m^{t+1}}{k_{0}}+\Omega_{4}\right)\left(\frac{1+\Omega_{0}}{k_{t+1}-\Omega_{1}}\right)\right]}$ and $\Psi_{2}=\frac{k_{t+1}+t-1}{k_{t+1}-\Omega_{1}}+\left(\frac{1+\Omega_{0}}{k_{t+1}-\Omega_{1}}\right)\Psi_{1}$, Eq.(\ref{eqa:MF-4-4-2-17}) may easily reduce to Eq.(\ref{eqa:MF-4-4-2-2}), implying that Theorem 4 is complete.

In terms of space limitation, we omit the related comparison between Eq.(\ref{eqa:MF-4-4-2-0}) and  Eq.(\ref{eqa:MF-4-4-2-1}). Interested reader can prove the equivalence between them in a similar manner mentioned in subsection 4.4.1.

\section{Conclusion}

In summary, we have presented a class of graphs $G^{\star}(m,t)$ with diameter $2$ and then considered some widely-studied structural properties on graphs $G^{\star}(m,t)$. Based on theoretical analysis and experimental simulations, we find that graphs $G^{\star}(m,t)$ show some intriguing structural properties including density feature, scale-free feature (degree distribution obeying power-law with exponent $2$ in form), small-world character and disassortative structure. At the same time, we have studied the trapping problem on graphs $G^{\star}(m,t)$ and precisely derived the closed-from solution to average trapping time. The results reveal that our graphs are able to have more optimal trapping efficiency by holding the theoretical lower bound for average tapping time.

It is worth noting that the goal of this work is to only study graphs of great interest from the theoretical point of view. In another words, all models considered may be artificial and rarely found in real-world situation. Yet, as shown above, we introduce a generative framework for creating scale-free graphs with exponent $2$, a class of models that are seldom addressed in early study. Furthermore, the methods developed for discussing the trapping problem of graphs may be useful to study many other models in the study of complex networks.

\section*{Acknowledge}

The research was supported in part by the National Key Research and Development Plan under grant 2017YFB1200704 and the National Natural Science Foundation of China under grant No. 61662066.


{\footnotesize
\begin{thebibliography}{9}

\setlength{\parskip}{0pt}


\bibitem{Newman-2018} M.E.J. Newman. Networks. Oxford university press. 2018

\bibitem{Albert-2016} A.-L. Barab\'{a}si. Network science. Cambridge university press. 2016

\bibitem{Shanker-2010} O. Shanker. Complex network dimension and path counts. Theoretical Computer Science, 411 (2010): 2454-2458

\bibitem{Hao-2018} L. Hao. Overlapping community detection with least replicas in complex networks. Information Sciences. 453 (2018): 216-226

\bibitem{Staudt-2016} C.L. Staudt, A. Sazonovs, H. Meyerhenke. NetworKit: A tool suite for large-scale complex network analysis. Network Science. 4 (2016): 508-530

\bibitem{Watts-1998} D.J. Watts, S.H. Strogatz. Collective dynamics of small-world networks. Nature. 393 (1998): 440-442

\bibitem{Albert-1999-1} A.-L. Barab\'{a}si, R. Albert. Emergence of Scaling in Random Networks. Science. 5439 (1999): 509-512

 \bibitem{Bondy-2008} J.A. Bondy. U.S.R. Murty. Graph Theory. Springer. 2008

 \bibitem{Sarma-2015} A.D. Sarma, A. R. Molla , G. Pandurangan. Distributed computation in dynamic networks via random
walks. Theoretical Computer Science. 581 (2015): 45-66

\bibitem{Colomer-de-Simon-2014} P. Colomer-de-Simon. M, Boguna. Double Percolation Phase Transition in Clustered Complex Networks. Phys. Rev. X. 4 (2014)041020

\bibitem{Ling-2019} S.Y. Ling, R.T. Xu, A.S. Bandeira. On the Landscape of Synchronization Networks: A Perspective from Nonconvex Optimization. SIAM Journal on Optimization. 29 (2019): 1879-1907



\bibitem{MF-2019-1} F. Ma, P. Wang, B. Yao. An extremal problem: How small scale-free graph can be. To submit. 	arXiv:1911.09253

\bibitem{Wang-2019} Y.C. Wang, Q. Bao, Z.Z. Zhang. Combinatorial properties of Farey graphs. Theoretical Computer Science. 796 (2019): 70-89

\bibitem{Ferretti-2017} S. Ferretti. On the modeling of musical solos as complex networks. Information Sciences. 375 (2017): 271-295

\bibitem{Ma-2018} F. Ma, B. Yao. An iteration method for computing the total number of spanning trees and its applications in graph theory. Theoretical Computer Science. 708 (2018): 46-57

\bibitem{Chen-2007} M. Chen, B.M. Yu, P. Xu, J. Chen. A new deterministic complex network model with
hierarchical structure. Physica A.  385 (2007): 707-717

\bibitem{Albert-2001} A.-L. Barab\'{a}si, E. Ravasz, T. Vicsek. Deterministic scale-free networks. Physica A. 299 (2001): 559-564

\bibitem{Ma-2019} F. Ma, P. Wang, B. Yao. Generating Fibonacci-model as evolution of networks with vertex-velocity and time-memory. Physica A. 527 (2019)121295

\bibitem{Ravasz-2003}  E. Ravasz,  A.-L. Barab\'{a}si. Hierarchical organization in complex networks. Phys. Rev. E. 67 (2003)026112


\bibitem{Definition-2020-2} Generally speaking, a graph $G(V,E)$ is defined as small-world if it has a smaller diameter $D$ and a higher clustering coefficient $\langle C\rangle$. Here, it is convention to require $D=O(\ln|V|)$ and $\langle C\rangle\rightarrow \alpha$ (a non-zero constant ) in the large graph size limit.

\bibitem{Newman-2002} M.E.J. Newman. Assortative mixing in networks. Phys. Rev. Lett. 89, (2002) 208701.

\bibitem{Rosvall-2008} M. Rosvall. C.T. Bergstrom. Maps of random walks on complex networks reveal community structure. Proc Natl Acad Sci U S A. 105 (2008): 1118-1123

\bibitem{Chung-1997} F.R. Chung. Spectral Graph Theory. American Mathematical Society. Providence. 1997


\bibitem{Aldous-1999} D. Aldous, J. Fill. Reversible Markov chains and random walks on graphs (1999), http://www.stat.berkeley.edu/aldous/RWG/Chap2.pdf

\bibitem{Zhang-2009} Z.Z. Zhang, Y. Qi, S.G. Zhou, Y. Lin, J.G. Guan. Recursive solutions for Laplacian spectra and eigenvectors of a class of growing treelike networks. Phys. Rev. E. 80 (2009)016104.

\bibitem{Benichoua-2014} O. Benichoua, R. Voituriez. From first-passage times of random walks in confinement to geometry-controlled kinetics. Physics Reports. 539 (2014): 225-284


\bibitem{Definition-2020-1} In general, it is a trivial fact that the scheme for assigning number $|V|-1$ of traps to all vertices but an arbitrary vertex in a given graph $G(V,E)$ must be perfect.

\bibitem{J. Stelling-2002} J. Stelling, S. Klamt, K. Bettenbrock, S. Schuster, E.D. Gilles. Nature. 420 (2002): 190-193

\bibitem{A. Akhmanova-2008} A. Akhmanova, M.O. Steinmetz. Nature Reviews Molecular Cell Biology. 9 (2008): 309-322





\end{thebibliography}
}
\end{document}